\documentclass[aps,prd,superscriptaddress,reprint,amsmath,amssymb,nofootinbib]{revtex4-1}

\usepackage{graphicx}
\usepackage{dcolumn}
\usepackage{bm}
\usepackage{hyperref}
\usepackage[caption=false]{subfig}
\usepackage{multirow}

\begin{document}
\title{Determining neutrino oscillation parameters from atmospheric muon neutrino disappearance with three years of IceCube DeepCore data}

\affiliation{III. Physikalisches Institut, RWTH Aachen University, D-52056 Aachen, Germany}
\affiliation{School of Chemistry \& Physics, University of Adelaide, Adelaide SA, 5005 Australia}
\affiliation{Dept.~of Physics and Astronomy, University of Alaska Anchorage, 3211 Providence Dr., Anchorage, AK 99508, USA}
\affiliation{CTSPS, Clark-Atlanta University, Atlanta, GA 30314, USA}
\affiliation{School of Physics and Center for Relativistic Astrophysics, Georgia Institute of Technology, Atlanta, GA 30332, USA}
\affiliation{Dept.~of Physics, Southern University, Baton Rouge, LA 70813, USA}
\affiliation{Dept.~of Physics, University of California, Berkeley, CA 94720, USA}
\affiliation{Lawrence Berkeley National Laboratory, Berkeley, CA 94720, USA}
\affiliation{Institut f\"ur Physik, Humboldt-Universit\"at zu Berlin, D-12489 Berlin, Germany}
\affiliation{Fakult\"at f\"ur Physik \& Astronomie, Ruhr-Universit\"at Bochum, D-44780 Bochum, Germany}
\affiliation{Physikalisches Institut, Universit\"at Bonn, Nussallee 12, D-53115 Bonn, Germany}
\affiliation{Universit\'e Libre de Bruxelles, Science Faculty CP230, B-1050 Brussels, Belgium}
\affiliation{Vrije Universiteit Brussel, Dienst ELEM, B-1050 Brussels, Belgium}
\affiliation{Dept.~of Physics, Chiba University, Chiba 263-8522, Japan}
\affiliation{Dept.~of Physics and Astronomy, University of Canterbury, Private Bag 4800, Christchurch, New Zealand}
\affiliation{Dept.~of Physics, University of Maryland, College Park, MD 20742, USA}
\affiliation{Dept.~of Physics and Center for Cosmology and Astro-Particle Physics, Ohio State University, Columbus, OH 43210, USA}
\affiliation{Dept.~of Astronomy, Ohio State University, Columbus, OH 43210, USA}
\affiliation{Niels Bohr Institute, University of Copenhagen, DK-2100 Copenhagen, Denmark}
\affiliation{Dept.~of Physics, TU Dortmund University, D-44221 Dortmund, Germany}
\affiliation{Dept.~of Physics, University of Alberta, Edmonton, Alberta, Canada T6G 2E1}
\affiliation{Erlangen Centre for Astroparticle Physics, Friedrich-Alexander-Universit\"at Erlangen-N\"urnberg, D-91058 Erlangen, Germany}
\affiliation{D\'epartement de physique nucl\'eaire et corpusculaire, Universit\'e de Gen\`eve, CH-1211 Gen\`eve, Switzerland}
\affiliation{Dept.~of Physics and Astronomy, University of Gent, B-9000 Gent, Belgium}
\affiliation{Dept.~of Physics and Astronomy, University of California, Irvine, CA 92697, USA}
\affiliation{Dept.~of Physics and Astronomy, University of Kansas, Lawrence, KS 66045, USA}
\affiliation{Dept.~of Astronomy, University of Wisconsin, Madison, WI 53706, USA}
\affiliation{Dept.~of Physics and Wisconsin IceCube Particle Astrophysics Center, University of Wisconsin, Madison, WI 53706, USA}
\affiliation{Institute of Physics, University of Mainz, Staudinger Weg 7, D-55099 Mainz, Germany}
\affiliation{Dept.~of Physics and Astronomy, Michigan State University, East Lansing, MI 48824, USA}
\affiliation{Universit\'e de Mons, 7000 Mons, Belgium}
\affiliation{Technische Universit\"at M\"unchen, D-85748 Garching, Germany}
\affiliation{Bartol Research Institute and Dept.~of Physics and Astronomy, University of Delaware, Newark, DE 19716, USA}
\affiliation{Dept.~of Physics, University of Oxford, 1 Keble Road, Oxford OX1 3NP, UK}
\affiliation{Dept.~of Physics, Drexel University, 3141 Chestnut Street, Philadelphia, PA 19104, USA}
\affiliation{Physics Department, South Dakota School of Mines and Technology, Rapid City, SD 57701, USA}
\affiliation{Dept.~of Physics, University of Wisconsin, River Falls, WI 54022, USA}
\affiliation{Oskar Klein Centre and Dept.~of Physics, Stockholm University, SE-10691 Stockholm, Sweden}
\affiliation{Dept.~of Physics and Astronomy, Stony Brook University, Stony Brook, NY 11794-3800, USA}
\affiliation{Dept.~of Physics, Sungkyunkwan University, Suwon 440-746, Korea}
\affiliation{Dept.~of Physics, University of Toronto, Toronto, Ontario, Canada, M5S 1A7}
\affiliation{Dept.~of Physics and Astronomy, University of Alabama, Tuscaloosa, AL 35487, USA}
\affiliation{Dept.~of Astronomy and Astrophysics, Pennsylvania State University, University Park, PA 16802, USA}
\affiliation{Dept.~of Physics, Pennsylvania State University, University Park, PA 16802, USA}
\affiliation{Dept.~of Physics and Astronomy, Uppsala University, Box 516, S-75120 Uppsala, Sweden}
\affiliation{Dept.~of Physics, University of Wuppertal, D-42119 Wuppertal, Germany}
\affiliation{DESY, D-15735 Zeuthen, Germany}

\author{M.~G.~Aartsen}
\affiliation{School of Chemistry \& Physics, University of Adelaide, Adelaide SA, 5005 Australia}
\author{M.~Ackermann}
\affiliation{DESY, D-15735 Zeuthen, Germany}
\author{J.~Adams}
\affiliation{Dept.~of Physics and Astronomy, University of Canterbury, Private Bag 4800, Christchurch, New Zealand}
\author{J.~A.~Aguilar}
\affiliation{D\'epartement de physique nucl\'eaire et corpusculaire, Universit\'e de Gen\`eve, CH-1211 Gen\`eve, Switzerland}
\author{M.~Ahlers}
\affiliation{Dept.~of Physics and Wisconsin IceCube Particle Astrophysics Center, University of Wisconsin, Madison, WI 53706, USA}
\author{M.~Ahrens}
\affiliation{Oskar Klein Centre and Dept.~of Physics, Stockholm University, SE-10691 Stockholm, Sweden}
\author{D.~Altmann}
\affiliation{Erlangen Centre for Astroparticle Physics, Friedrich-Alexander-Universit\"at Erlangen-N\"urnberg, D-91058 Erlangen, Germany}
\author{T.~Anderson}
\affiliation{Dept.~of Physics, Pennsylvania State University, University Park, PA 16802, USA}
\author{C.~Arguelles}
\affiliation{Dept.~of Physics and Wisconsin IceCube Particle Astrophysics Center, University of Wisconsin, Madison, WI 53706, USA}
\author{T.~C.~Arlen}
\affiliation{Dept.~of Physics, Pennsylvania State University, University Park, PA 16802, USA}
% ...

\date{\today}
\collaboration{IceCube Collaboration}
\noaffiliation

\date{\today}
\begin{abstract}
We present a measurement of neutrino oscillations via atmospheric muon neutrino disappearance with three years of data of the completed IceCube neutrino detector. DeepCore, a region of denser IceCube instrumentation, enables the detection and reconstruction of atmospheric muon neutrinos between 10\,GeV and 100\,GeV, where a strong disappearance signal is expected. The IceCube detector volume surrounding DeepCore is used as a veto region to suppress the atmospheric muon background. Neutrino events are selected where the detected Cherenkov photons of the secondary particles minimally scatter, and the neutrino energy and arrival direction are reconstructed. Both variables are used to obtain the neutrino oscillation parameters from the data, with the best fit given by $\Delta m^2_{32}=2.72^{+0.19}_{-0.20}\times 10^{-3}$\,eV$^2$ and $\sin^2\theta_{23} = 0.53^{+0.09}_{-0.12}$ (normal mass ordering assumed). The results are compatible, and comparable in precision, to those of dedicated oscillation experiments.
\end{abstract}
\maketitle

In the 1990s, Super-Kamiokande's measurements of atmospheric neutrinos \cite{sk1} led to the acceptance of the mass-induced oscillation model where neutrinos are massive particles whose interaction eigenstates do not have an exact correspondence to their mass eigenstates. This property gives neutrinos produced in one flavor eigenstate, $\alpha$, a non-zero probability of interacting in a different flavor, $\beta$, after traveling for some distance $L$. In the simplest scenario, with only two neutrino flavors, the transition probability is given by
\begin{equation}
  \label{eq:osc_prob}
  P(\nu_\alpha \rightarrow \nu_\beta) = \sin^2 \left(2 \theta\right)\,\sin^2\left(\Delta m^2\,\frac{L}{4\,E_\nu}\right),
\end{equation}
where $\theta$ defines the mixing between mass and flavor eigenstates, $\Delta m^2$ is the difference in the squared masses and $E_\nu$ is the neutrino energy (in natural units). Considering the existence of three neutrinos, as done in this letter, provides an oscillation probability that consists of a sum of terms of the form of Eq.\,\ref{eq:osc_prob}, but involving three mixing angles, two mass-squared differences and a complex phase.

Currently, the mixing angles, the solar mass splitting and the absolute value of the atmospheric mass splitting have been measured \cite{kamland_results, sno_results, borexino_results, dayabay_results, reno_results, dchooz_results, t2k_disappearance, minos1, sk3} while the existence of CP-violation and the ordering of the masses remain open questions \cite{osc_status1, osc_status2}. Addressing these questions requires improving the measurement precision on the known parameters and improving the measurements sensitive to the parameters that modify the oscillation probabilities as neutrinos traverse matter \cite{oscillations_inice, future_akhmedov, future_winter}.

In the following, we focus on the measurement of the oscillation parameters $\theta_{23}$ and $\Delta m^2_{32}$. The measurement presented here achieves a precision comparable to the leading experiments in the field \cite{minos1,t2k_disappearance,sk3} using a sample of atmospheric high energy neutrinos, from 10\,GeV to 100\,GeV, recorded with a sparsely instrumented detector located in a natural medium.

\section{The IceCube detector} 
The data used in this analysis were collected by IceCube \cite{icecube}, an ice Cherenkov neutrino detector located at depths between 1450\,m and 2450\,m at the geographic South Pole. IceCube consists of 5160 downward-facing 10-inch PMTs, enclosed in glass pressure spheres, known as digital optical modules (DOMs) \cite{icecube_daq}. 

The detector is an array of 86 strings, each holding 60 DOMs. Of these, 78 strings are arranged in a triangular grid with a typical distance of 125\,m between the strings and a vertical distance of 17\,m between DOMs on the strings. The lower center region of the array, from 1760\,m down to 2450\,m, houses a region of denser instrumentation (7\,m DOM spacing) known as DeepCore \cite{dcdesign} with eight strings at string-to-string distances between $40-70$\,m. Some 50\,\% of the PMTs in this region have 35\,\% higher quantum efficiency than the standard IceCube PMTs. The result is a neutrino energy threshold in DeepCore an order of magnitude smaller than in IceCube, of about 10\,GeV.

\section{Neutrino oscillations in IceCube DeepCore} 
The IceCube detector records more than $10^5$ atmospheric neutrinos every year, a large fraction of them in the DeepCore sub-array \cite{dcdesign}. These neutrinos cover path lengths through the Earth ranging from 10\,km to about 12700\,km depending on their arrival zenith angle, $\theta_z$. Above GeV energies they follow a steeply falling spectrum \cite{frejus_spectrum, sk_spectrum, amanda_spectrum, antares_spectrum, icecube_unfolding} that covers several orders of magnitude up to a few hundred TeV.

Neutrino oscillations modify the flavor ratio of the flux of atmospheric neutrinos measured at the detector site. The strongest effect to which DeepCore is sensitive is the disappearance of $\nu_\mu$, modulated by the large (atmospheric) mass splitting, with $\Delta m_{32}^2 \simeq \Delta m_{31}^2$, and the mixing angle $\theta_{23}$ \cite{oscillations_inice}. In this analysis, these parameters are derived from the distortions on the expected $\nu_\mu$ flux.

 %The best measured values of these parameters \cite{t2k_disappearance, minos1, sk3} predict a maximum disappearance at $E_\nu \sim 25\,$GeV for paths that cross the entire Earth $(\cos\theta_z = -1$). Neutrinos with lower energies reach maximum disappearance in shorter baselines $(\cos\theta_z > -1)$.

\begin{figure}[bt!]
  \subfloat[\label{fig:event_view_a}Event side view.]{
    \includegraphics[trim= 10 0 10 0,width=0.195\textwidth]{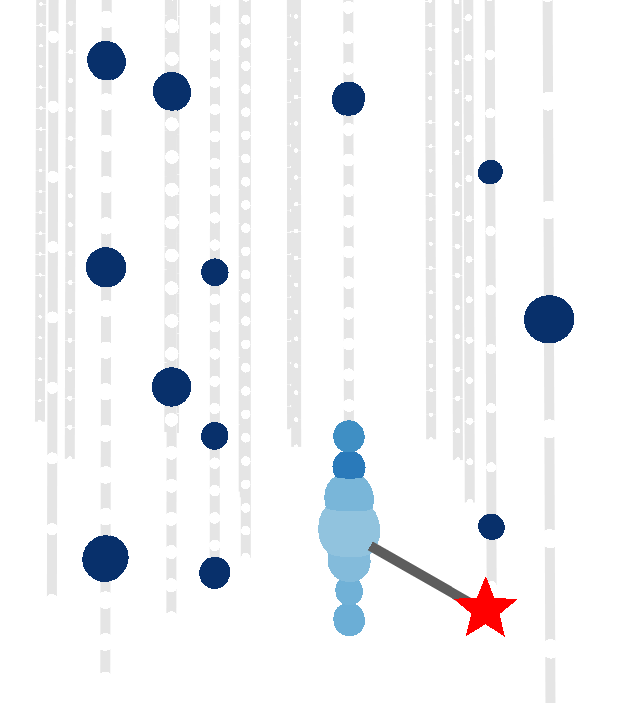}
    }
  \subfloat[\label{fig:event_view_b}Projection onto one string.]{
    \includegraphics[trim = 0 10 0 0, clip,width=0.255\textwidth]{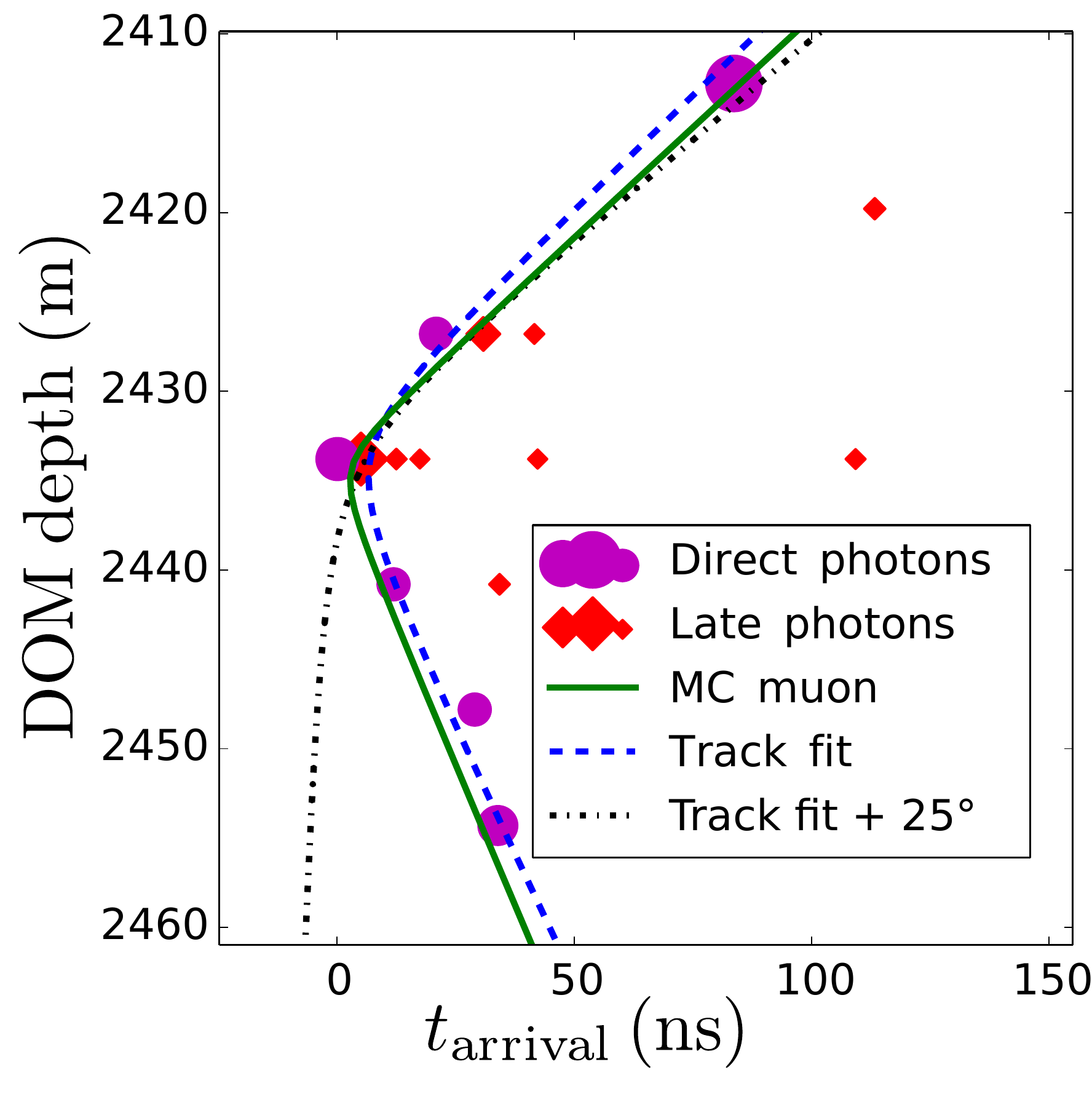}
    }
  \caption{\label{fig:event_view}A simulated 12\,GeV $\nu_\mu$ interacting in DeepCore and producing an 8\,GeV muon (42\,m range) and a 4\,GeV hadronic shower. In (a), the dashed vertical lines are detector strings, the star marks the position of the interaction vertex and the solid line is the muon track. Twenty DOMs record photons and they have colors related to the photon arrival time (lighter is earlier) where their size is proportional to the charge observed. In (b), the DOM depth as a function of the arrival time of photons at the string with most light collected is shown. Marker sizes scale with charge. The expected hyperbolae from simulation, a track fit and the same fit altered by 25$^\circ$ are also shown.}
\end{figure}

Muon neutrino charged current (CC) interactions in the ice with energies between 10\,GeV and 100\,GeV, the primary signal event in this analysis, typically produce a minimum-ionizing muon track and initiate a hadronic shower, both of which emit Cherenkov light. The signature of these interactions in DeepCore are individual Cherenkov photons that are partially scattered due to the optical properties of the ice. Figure\,\ref{fig:event_view} shows the detector's response for one such interaction.

\begin{figure*}[th]
 \includegraphics[trim=0 10 0 0,width=\textwidth]{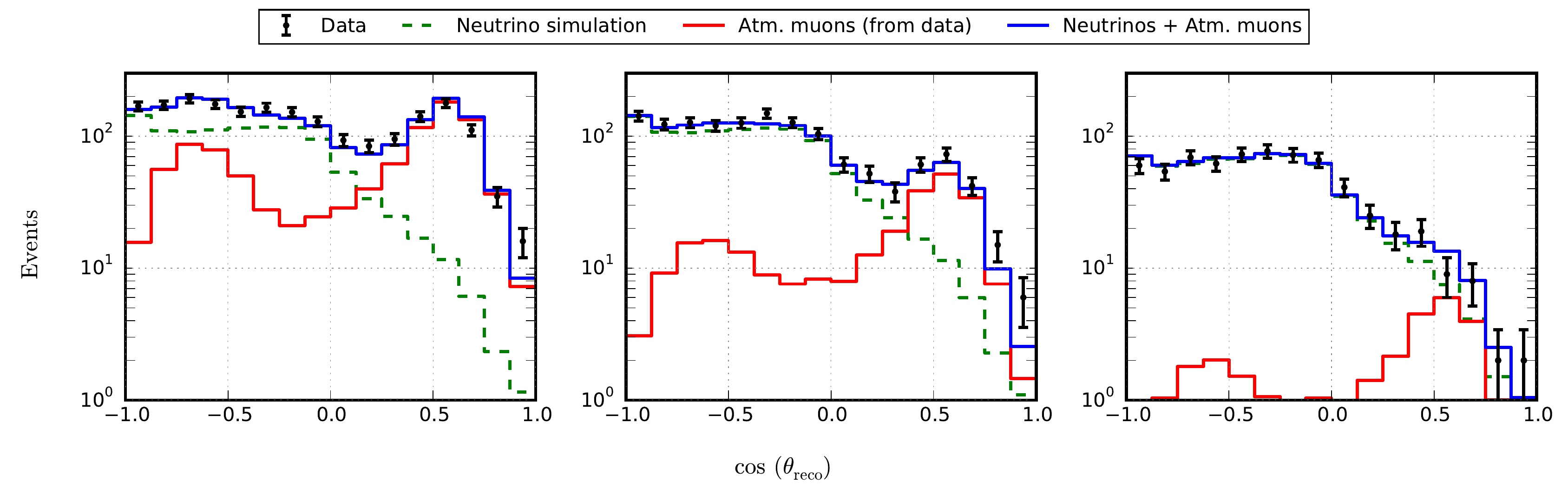}
 \caption{\label{fig:muon_background}Zenith angle distributions of neutrino simulation and atmospheric muons derived from data for three subsequent steps in the event selection with increasing veto cuts. To go from the first to the second panel the veto cut which uses a muon track hypothesis is applied. A cut on the charge observed above and prior to the trigger is used to go from the second to the third panel. A comparison is also made to a 10\,\% control sample of the data. The small excess in the data around $\cos\theta_z \simeq 0.3$ in the first panel are atmospheric muons that could not be tagged. Note that the region $\cos\theta_z > 0$ is not used in the final analysis of the data.}
\end{figure*}

The dominant sources of background for this measurement are muons from cosmic ray showers, CC interactions of electron and tau neutrinos, and neutral current interactions of all neutrino flavors. Atmospheric muons trigger the detector at a rate $10^5$ times higher than the $\nu_\mu$ CC rate, which itself is three times higher than the combined rate of all other neutrino interactions. 

Muons from cosmic ray showers are generated using the CORSIKA package \cite{corsika}. The atmospheric neutrino simulation follows the flux predicted by Honda et al \cite{honda12}, while neutrino-nucleon interactions in ice are simulated using the GENIE software \cite{genie}. Muons are propagated according to the parameterization presented in \cite{mmc}, while all other particles are passed to the Geant4 package \cite{geant4}.

\section{Event selection} 
The event selection, described in detail in \cite{yanez_thesis}, has the goal of identifying events that start in the detector volume with a clear muon track to reduce the background. To avoid contamination from atmospheric muons, the data analyzed consists only of events reconstructed as passing through the Earth ($\cos\theta_z \leq 0 $). However, while atmospheric muons enter the detector only from above ($\cos\theta_z \geq 0 $), the small probability of a mis-reconstruction combined with the large number of events detected results in a significant pollution of the neutrino sample. The event selection starts by rejecting atmospheric muons using the dedicated DeepCore trigger and filter \cite{dcdesign}.

\subsection{Rejection of atmospheric muon background}
The atmospheric muon background which remains after the DeepCore filter is removed by searching for muon tracks that enter the DeepCore volume from outside and pass near the DOMs that triggered the detector. This uses the outer part of the DeepCore detector as an effective veto region, similar to that described in \cite{ic79_cascades}. Atmospheric muon simulation is used to understand the basic characteristics of the background and develop methods to remove it. The statistics available are, however, not enough to provide a complete description of the background at the final level of the analysis and detector data is used instead.

In this analysis, cuts on the position of the earliest DOM involved in the trigger, the total charge observed in the DOMs above and prior to the trigger, the charge collected as a function of time $(dQ/dt)$, and the number of DOMs above threshold in a narrow time window $\left[-150\,\mathrm{ns},+250\,\mathrm{ns}\right]$ in coincidence with the photons expected from an atmospheric muon hypothesis are applied. Events reconstructed with $\cos \theta_z > 0$ by a fast track reconstruction algorithm \cite{improved_linefit} and a maximum likelihood reconstruction \cite{amanda_reco} are also tagged as atmospheric muons. The veto selection cuts reduce the atmospheric muons to similar rates as the neutrino events while keeping about 40\,\% of the original muon neutrino sample.

The last veto method listed, which uses a muon track hypothesis, is particularly sensitive to muons which enter the fiducial volume through the corridors formed by the detector geometry, leaving very little detected light. The number of photons observed in an event therefore depends primarily on its azimuth arrival direction, and is largely independent of the event characteristics inside the fiducial volume (with variations of less than 10\,\%). By selecting events above the noise threshold of the search, a sample of atmospheric muons which fulfill the quality criteria, outlined next, is obtained. These events are used to create the template for the muon background at the final selection level. 

Figure\,\ref{fig:muon_background} shows the zenith angle distribution of a subsample of the data at three steps in the selection process, where the contributions from neutrinos and atmospheric muons are given separately. In the figure, the region where $\cos \theta_z \geq 0$ is also shown, even though it is not used for obtaining the final result. The transition from a muon to neutrino (assuming the best known oscillation parameters) dominated sample as additional selection criteria are applied can be seen in the three steps. The normalization of both components is fit for each figure, and the results are consistent with those obtained from the fit of the oscillation parameters.

\subsection{Selection and reconstruction of $\nu_\mu$ interactions}
The neutrino interactions of interest result in a small number of DOMs with photon hits (see Fig.\,\ref{fig:event_view_a}).  It is likely that most photons will have scattered before detection and requiring a minimum number of \textit{direct} photons preferentially selects events that occur close to a string. This reduces the impact of optical scattering in the ice and ensures a well reconstructed event. 

Direct photons are identified by exploiting the fact that Cherenkov light is emitted at a fixed angle relative to the direction of the charge particle and thus the depth at which a photon arrives at a DOM on a string as a function of its time of arrival is described by a hyperbola (see Fig.\,\ref{fig:event_view_b} and \cite{bbfit}). When photons scatter they follow a longer path length, resulting in a delay that makes them fall outside the expected hyperbola. Direct photons from multiple Cherenkov emitters are also well approximated by a single hyperbola.

The direct photon identification procedure looks for signals that match a hyperbolic pattern while iterating over one string at a time. In such a search there is no need to assume a track or hadronic shower hypothesis. Each DOM is characterized by the time of arrival of the earliest photon in the event and the total observed charge. The DOM with the highest charge is used as the starting point. A time window is defined for accepting a photon in the DOM directly above or below, given by $|\Delta z_\textrm{DOMs}/c_\mathrm{ice}| \pm t_\mathrm{delay}$, where $\Delta z_\textrm{DOMs}$ is the distance between the DOMs, $c_\textrm{ice}$ is the speed of light in ice and $t_\mathrm{delay}$ is the permitted time delay, set to 20\,ns for this analysis. The selected DOMs of a given string are considered directly hit only if three or more are found.

\begin{figure}[bt!]
 \includegraphics[width=0.95\columnwidth]{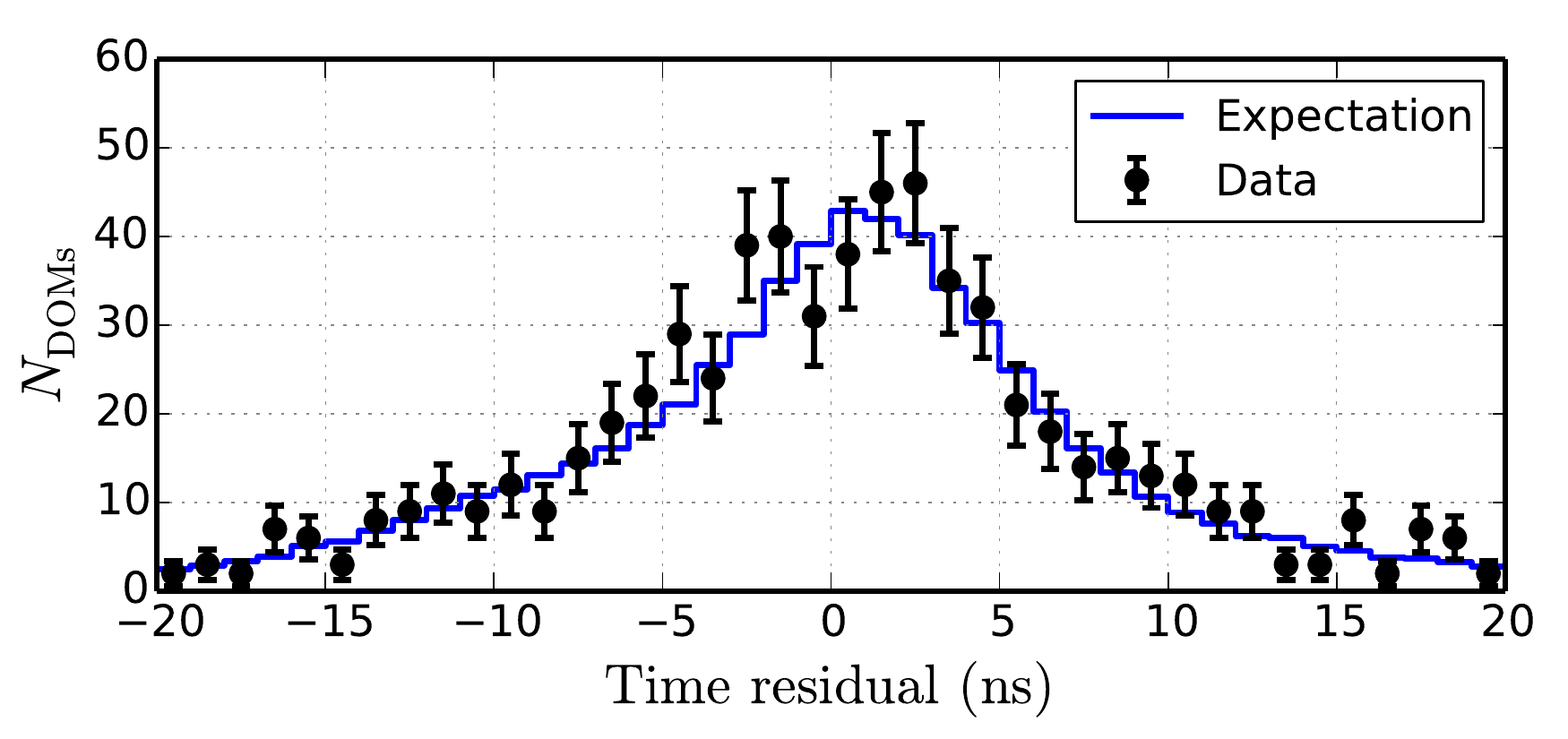}
 \caption{\label{fig:pulse_delay_exp}Difference in arrival time between direct photons and the expected hyperbola from the track fit, comparing simulation with 5\,\% of the final data sample.}
\end{figure}

Direct photons identified by this method have a mean arrival time delay, due to minimal scattering, of 18\,ns, compared to a typical mean delay time of 230\,ns. An event is selected for subsequent processing if at least a total of five DOMs with direct photons are found. This keeps about $30\,\%$ of the muon neutrinos in the relevant energy and zenith angle range. The agreement between data and simulation after this cut is shown in Fig.\,\ref{fig:pulse_delay_exp}.

Following \cite{bbfit}, the direct photons of an event are used to fit two topology hypotheses for Cherenkov emission, a single point (hadronic shower) and along a track (muon), using a $\chi^2$ optimization where no scattering is assumed. The $\chi^2$ ratio of the track-like and point-like hypotheses is used to select events with a muon track. Requiring a track assures that the event has directional information but retains only 30\,\% of the $\nu_\mu$ CC interactions with direct photons, as not all interactions produce a sufficiently long (20\,m) muon track. The zenith angle obtained from the fitted track hypothesis is used as one of the observables in the measurement, and the reduced-$\chi^2$ of the fit is used as a cut variable. 

The method used for estimating the total energy of a neutrino event makes the assumption that all interactions produce a hadronic shower at the interaction vertex, the brightness of which scales with energy, and a minimum-ionizing muon, assuming constant energy loss, that are emitted in the same direction. The total neutrino energy is then determined by the range of the produced muon $R_\mathrm{muon}$ and the energy of the hadronic shower $E_\mathrm{shower}$,
\begin{equation}
E_\nu \simeq E_\mathrm{shower} + a\,R_\mathrm{muon},
\label{eq:full_energy}
\end{equation}
where $a$ is the constant energy loss for muons (in ice $a = 0.226$\,GeV/m).

The directions of the muon and the hadronic shower are held fixed in the reconstruction. Expectations for light from the tracks and electromagnetic showers at any given DOM are obtained from a multi-dimensional parameterization of many different source configurations, as explained and used in \cite{icecube_energy}. The light expected from a hadronic shower is obtained by scaling-down the expectation from an electromagnetic shower. Unlike  the directional fit, here both direct and scattered photons are used to reconstruct the neutrino energy.

The energy estimation is completed in two steps. The goal of the first step is to determine the range of the muon by assuming all the light present in the detector is explained by a single muon. The vertex and decay point of the muon are fit by maximizing the likelihood of a finite muon track, to explain the pattern observed, normalized to the probability obtained from an infinite muon track hypothesis.

The fit starts by finding the projection of the first and last DOM with a signal along the track, serving as first guesses for the vertex and decay point. Two independent likelihood functions, which are formed by multiplying the probability for a DOM to observe zero photons given an expectation of $x$ photons from a track, $P(0|x_\mathrm{track})=e^{-x}$, are maximized for each of these points. The DOMs considered are those that observed no light and are situated before the first guess vertex in one likelihood, and after the first guess decay point in the second (within a distance of 200\,m from the infinite track hypothesis). DOMs that detected light are not included in the calculation. The likelihood functions are given by
\begin{equation}
%\mathcal{L} = \dfrac{\mathcal{L}(finite)}{\mathcal{infinite}} = \dfrac{P(\mathrm{no-hit|finite track})}{P\mathrm{no-hit|infinite track}}
%\mathcal{L}_\mathrm{track} = \dfrac{\mathcal{L}(\mathrm{finite})}{\mathcal{L}(\mathrm{infinite})} = \prod_i \dfrac{\exp(-x_{i,\mathrm{finite}})}{\exp(-x_{i,\mathrm{infinite}})},
\mathcal{L}(0;x_\mathrm{track}) = \prod_j \dfrac{\exp(-x_{j,\mathrm{finite}})}{\exp(-x_{j,\mathrm{infinite}})},
\label{eq:llh_finitereco}
\end{equation}
where $j$ runs over all DOMs that fulfill the criteria outlined above for the vertex and decay point. This procedure has a typical accuracy of 25\,m.

The vertex point found in the first step above is used as a seed in the second step of the energy reconstruction where the aim is to describe the light in the vicinity of the interaction vertex taking into account that a hadronic shower might have also been produced. While the light output of the muon is taken to be constant along its range, the light expected from the hadronic shower depends on its energy. Both the energy of the hadronic shower and the position of the interaction vertex are obtained by maximizing a likelihood function similar to that in Eq.\,\ref{eq:llh_finitereco}, but which also includes DOMs that have detected light, a probability given by $P(1|x_\mathrm{vertex})=1-e^{-x}$. The likelihood function for the energy deposited at the vertex is then
\begin{equation}
\mathcal{L}(0/1;x_\mathrm{vertex}) =  \prod_i \left[1 - \exp(-x_i)\right] \prod_j \exp(-x_j),
\label{eq:llh_leera}
\end{equation}
where $x$ is the sum of the light expectation from a muon track, a hadronic shower and noise, and the subscripts $i$ and $j$ run over the DOMs within  a 300\,m radius of the vertex which have observed some or no light, respectively. The energy is finally obtained from evaluating Eq.\,\ref{eq:full_energy} with the decay point of the muon fit in the first step and the vertex position and hadronic shower energy obtained in the second step.

%% To estimate the neutrino energy, events are assumed to be $\nu_\mu$ CC interactions, composed of a single hadronic shower and a minimum-ionizing muon track originating at the interaction point, starting inside the DeepCore volume. Two likelihoods are used to evaluate the probability for observing light from these sources at the DOMs in the vicinity of the reconstructed track direction. The likelihoods are a function of the mean expected number of photons, ignoring time of arrival, which depends on the position of the interaction point along the fit direction, the brightness of the hadronic shower and the range of the muon. Both direct and scattered photons are used (see \cite{supplemental_section} for more details).

\begin{figure}[bt!]
 \includegraphics[width=0.95\columnwidth]{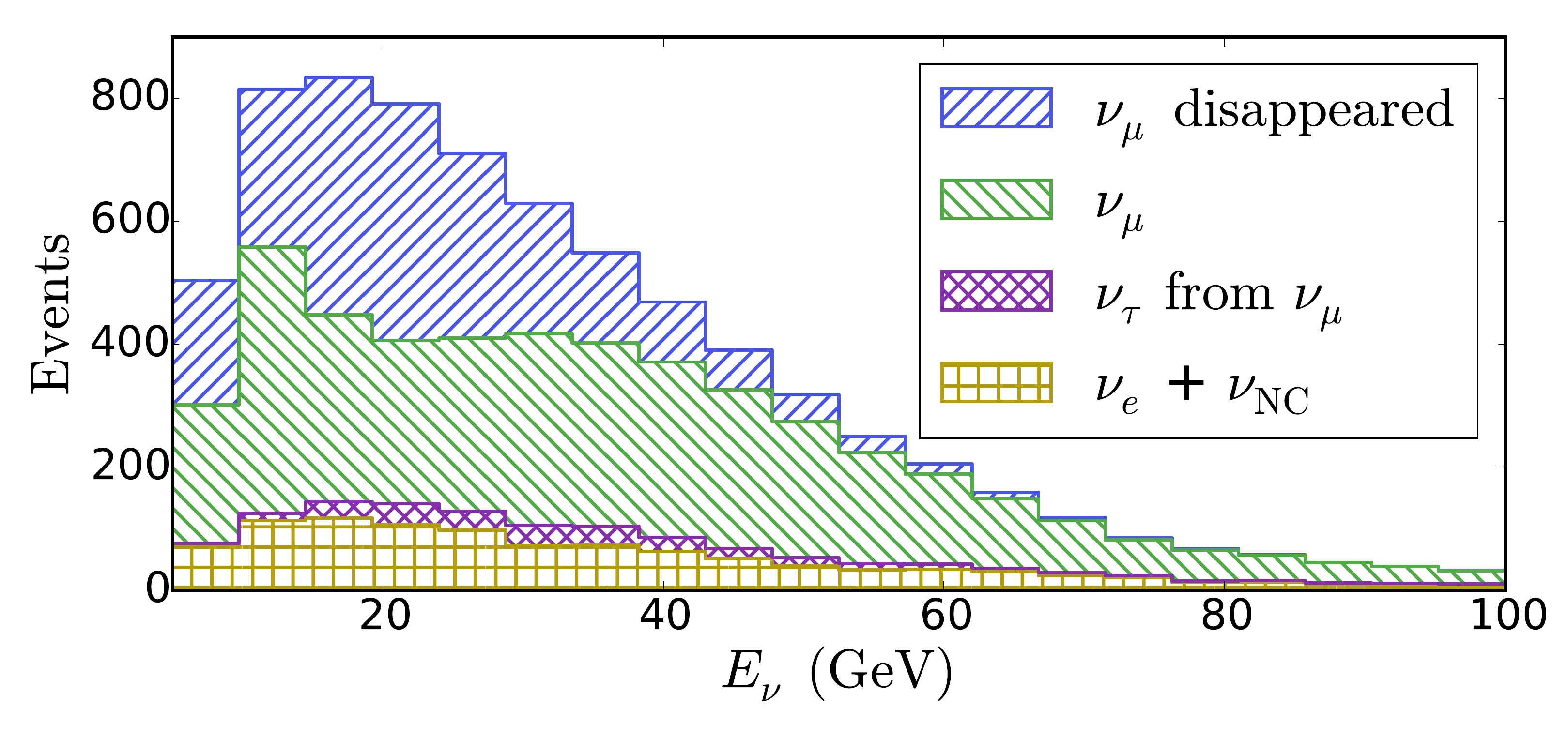}
 \caption{\label{fig:sample_energy}True energy distribution of simulated neutrino events in the final sample, with hatched areas representing each component. Only events used for the final result are considered ($E_\mathrm{reco} = [6,56]\,$GeV, and $\cos(\theta_\mathrm{reco}) < 0$). The missing $\nu_\mu$ component is also shown.}
\end{figure}

The energy distribution of the final analyzed neutrino sample, as given by the simulation, is shown in Fig.\,\ref{fig:sample_energy}. The sample\footnote{A detailed description of the final event selection can be found in \url{http://icecube.wisc.edu/science/data}.} is composed of 74\,\% $\nu_\mu$ CC, 13\,\% $\nu_e$ CC, 8\,\% neutral current interactions and 5\,\% $\nu_\tau$ CC. Deep inelastic scattering (DIS) events constitute 80\,\% of the sample, followed by resonant and quasi-elastic interactions. The expected atmospheric muon contribution to the final sample, from simulation, is less than 5\,\%. The median zenith angle resolution obtained for $\nu_\mu$ events is 12$^\circ$ at $E_\nu=10$\,GeV and improves to 5$^\circ$ at $E_\nu=40$\,GeV. The median energy resolution is 30\,\% at 8\,GeV and improves to 20\,\% at 15\,GeV.

\section{Data analysis} 
To determine the oscillation parameters, a binned maximum likelihood is used that includes nuisance parameters to account for systematic uncertainties \cite{pdg2014}. The data are binned in an 8$\times$8 two-dimensional histogram as a function of $\log_{10}(E_\mathrm{reco}/\mathrm{GeV})$, between 6\,GeV and 56\,GeV, and $\cos(\theta_\mathrm{reco})$ between -1 and 0. The physics parameters of the fit are the mixing angle $\theta_{23}$ and the mass splitting $\Delta m^2_{32}$. Oscillation probabilities are calculated using a full three-flavor scheme \cite{barger_osc, prob3}, including the effects of the Earth's matter distribution \cite{prem}. The mixing angle $\theta_{13}$ is treated as a nuisance parameter using the constraints from \cite{pdg2014}. The remaining oscillation parameters are fixed to the values given in \cite{fogliOsc}, as their uncertainties have a negligible impact on the result. 

The atmospheric neutrino flux uncertainties, including the normalization, varied by 20\,\%, the spectral index, with an uncertainty of 0.04, and the ratio of $\nu_e$/$\nu_\mu$, varied by 20\,\%, are considered. The template for the atmospheric muon contribution is obtained from inverting the veto cuts on the data, and its normalization is unconstrained in the fit.

Cross-section uncertainties are estimated from \cite{genie,gazizov,sarkar}. For DIS events the total cross-section is modified by 5\,\% and its energy dependence is changed by $E^{\pm0.03}$. Both uncertainties are smaller than the atmospheric neutrino flux uncertainties which impact the expectation in the same way. Non-linear energy dependencies occur at $E_\nu \leq 7$\,GeV. With less than 3\,\% of the sample below that threshold, the impact was found to be negligible. The effects of the uncertainty on resonant and quasi-elastic axial masses on the cross-section of non-DIS events, of $\pm 20\,\%$ and ${}_{-15\,\%}^{+25\,\%}$ respectively, were tested and found to modify the result by less than 1\,\%. The energy scale of hadronic showers, a 5\,\% bias in the energy estimation for MINOS \cite{minos_hadronic_scale, minos_result_hadronic_scale}, was assessed by varying the reconstructed hadronic energy, known to a resolution of 30\,\%, and was found to have no impact on the result.

The optical efficiency, or energy scale, of the DOMs has been calibrated in-situ using LEDs integrated into the DOMs and minimum-ionizing muons, and is known with an accuracy better than 10\,\% \cite{icecube_energy}. The angular acceptance of the DOMs, defined by the properties of the surrounding refrozen ice columns, is obtained from fitting the LED data and its uncertainty is estimated to be between 30\,\% and 10\,\%, depending on the incident angle. The impact of the description of the optical properties of the pristine glacial ice is tested by analyzing the data with two independent models \cite{ice_mie, ice_lea}.

Proper inclusion of the effect of uncertainties related to the detection process in the minimization proccedure would require events to be re-simulated with the modified values at each step. Full re-simulation of the sample at each minimization step is too computationally intensive, thus an alternative method has been introduced. 

The most important detector related systematics are the optical efficiency, or energy scale, of the DOMs and the optical scattering in the ice columns where the DOMs have been deployed.  The latter modifies the DOM angular acceptance and both are parametrized as a function of a single variable. Complete simulation sets were produced for several values of each parameter within the expected uncertainties. The sets are passed through the same selection and reconstruction steps as the data. When a fit is performed, the different sets are used to fill the bins of a histogram used to compare data and simulation. 

For each bin, the deviation in the number of expected events with respect to the nominal value is parameterized as a linear function of the quantity varied, e.g. the optical efficiency. The bin expectation for a set of physics and nuisance parameters is then given by weighting the standard simulation, creating the two-dimensional histogram in energy and direction, and multiplying the bin content by the variation expected for the detector parameters demanded. The re-weighting scheme at the histogram level was found to succesfully reproduce simulation generated with modified detector settings, and is used in the fit to the data.

\section{Results and conclusions} 
The analysis was applied to the data taken between May 2011 and April 2014, corresponding to 953 days of detector livetime. A total of 5174 events were observed, compared to an expectation of 6830 events assuming no oscillations. The parameters that describe the data best are, for a normal mass ordering, $\sin^2\theta_{23} = 0.53^{+0.09}_{-0.12}$ and $\Delta m^2_{32}=2.72^{+0.19}_{-0.20}\times 10^{-3}$\,eV$^2$, and for an inverted mass ordering, $\sin^2\theta_{23} = 0.51^{+0.09}_{-0.11}$ and $\Delta m^2_{32}=-2.73^{+0.18}_{-0.21}\times 10^{-3}$\,eV$^2$.  There is no significant preference found for the mass ordering. The errors solely due to statistical uncertainties are ${}^{+0.06}_{-0.08}$ for $\sin^2\theta_{23}$, and ${}^{+0.14}_{-0.15}\times 10^{-3}$\,eV$^2$ for $\Delta m^2_{32}$. Statistical and systematic uncertainties have an almost equal contribution to the errors of the final result.

The two ice models tested return best-fit points which differ by 0.04 in $\sin^2\theta_{23}$ and by $0.02\times 10^{-3}$\,eV$^2$ in $\Delta m^2_{32}$. The model that yields the most conservative errors, 30\,\% larger in $\sin^2\theta_{23}$ and about 7\,\% larger in $\Delta m^2_{32}$, is taken for the final reported results. For both models, the values of the nuisance parameters at the best fit point are within the assumed uncertainties and the atmospheric muon contribution to the final sample is fit to 1\,\%, consistent with the expectation obtained from simulation. A cross check was performed by fitting the data in an unbounded two-flavor scheme, which yielded consistent results.

\begin{figure*}[bth!]
 \includegraphics[trim= 0 25 0 0,width=0.76\textwidth]{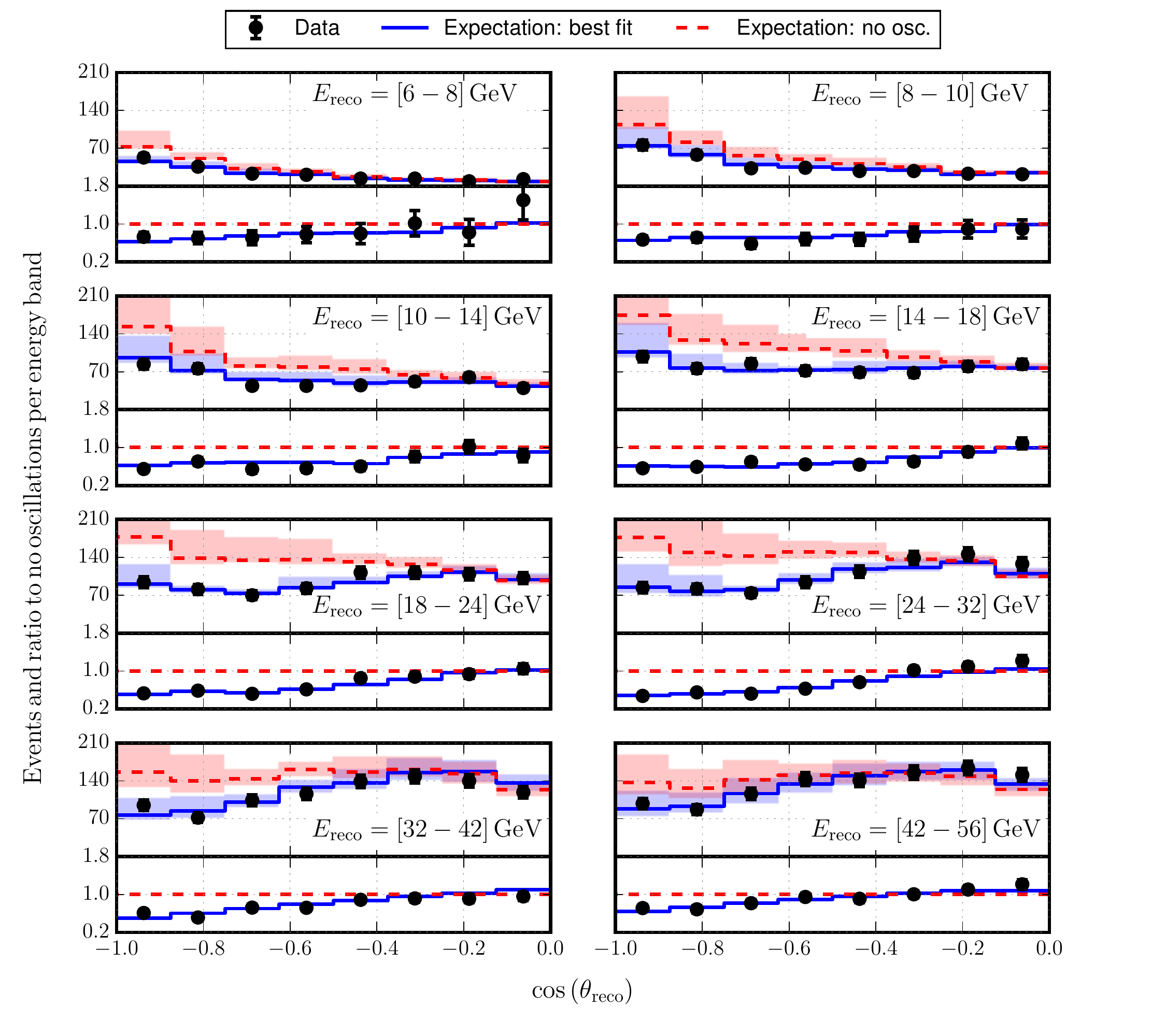}
 \caption{\label{fig:full_histogram}Comparison between data and expectations for the case of oscillations and no-oscillations. In each figure the zenith distribution for an energy band is shown (top), and the ratio of the data and the best-fit to no-oscillations is shown (bottom). The binning corresponds to that used for obtaining the best-fit oscillation parameters. Bands indicate the impact of the assumed systematic uncertainties.}
\end{figure*}
\begin{figure}[th!]
 \includegraphics[trim = 0 21 0 10, width=0.9\columnwidth]{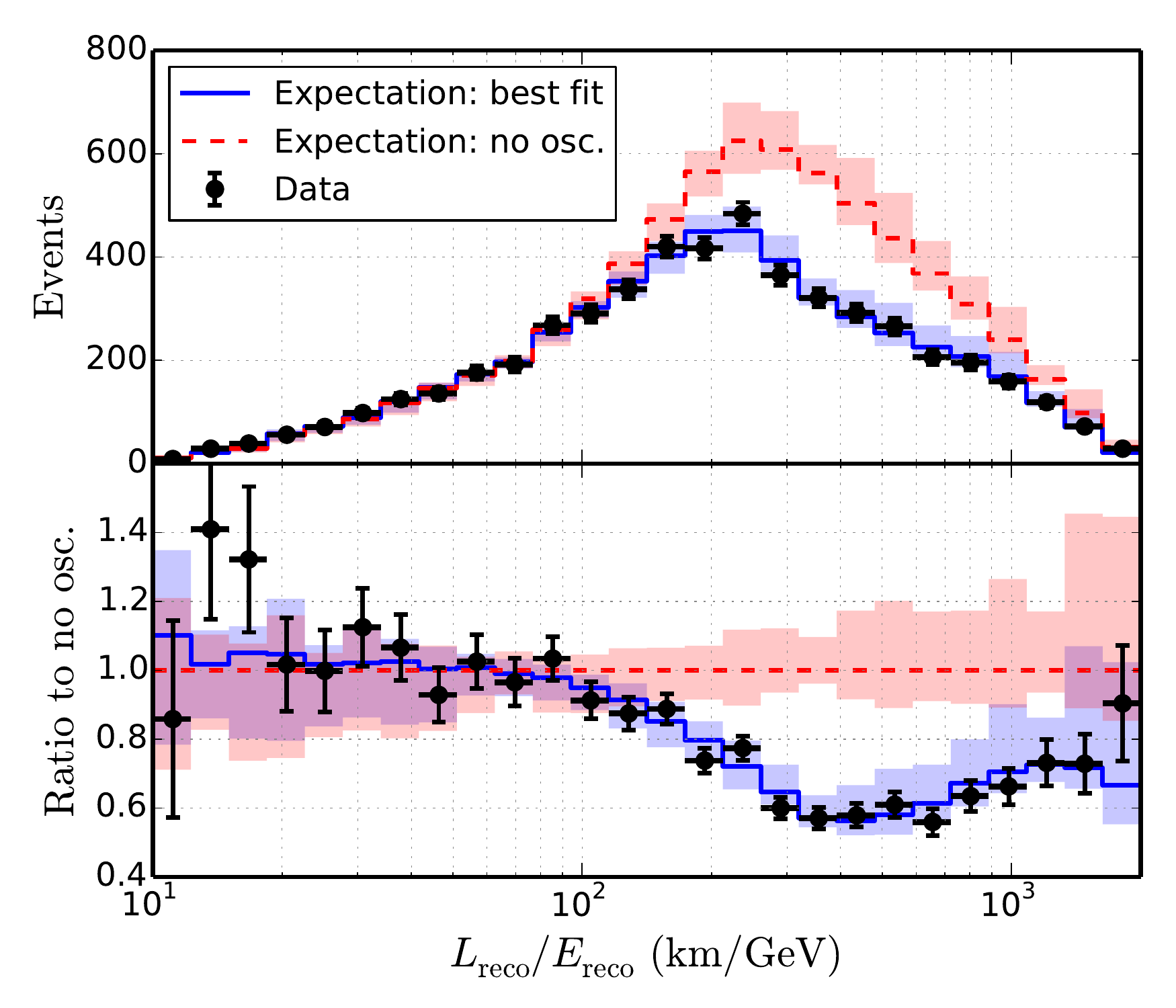}
 \caption{\label{fig:lefit}Distribution of events as a function of reconstructed $L/E$. Data are compared to the best fit and expectation with no oscillations (top) and the ratio of data and best fit to the expectation without oscillations is also shown (bottom). Bands indicate estimated systematic uncertainties.}
\end{figure}
\begin{figure}[tb!]
 \includegraphics[trim = 0 15 0 0 , width=0.99\columnwidth]{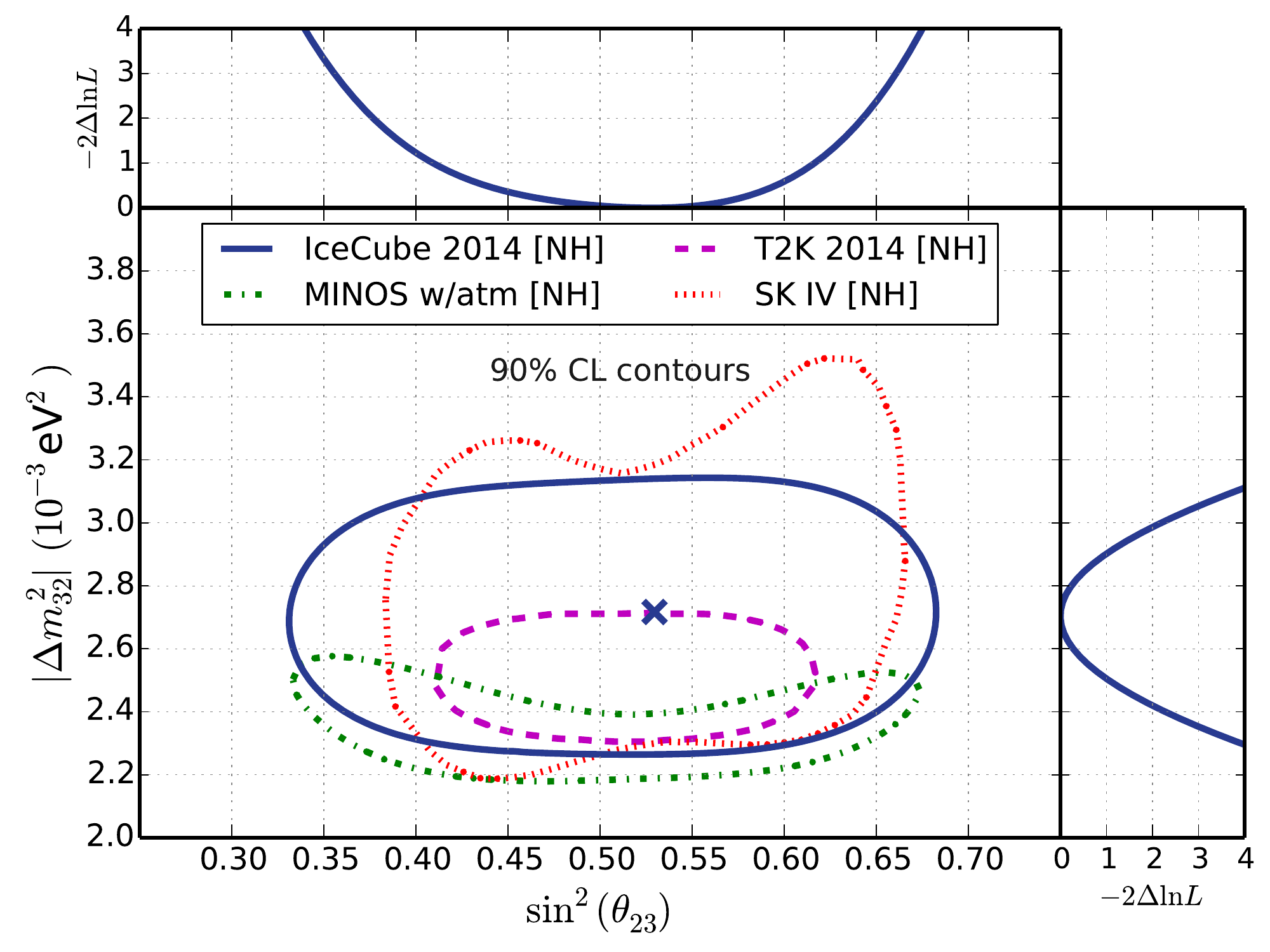}
 \caption{\label{fig:contours}90\,\% confidence contours of the result in the $\sin^2\theta_{23}-\Delta m^2_{32}$ plane in comparison with the ones of the most sensitive experiments \cite{t2k_disappearance, minos1, sk3}. The log-likelihood profiles for individual oscillation parameters are also shown (right and top). A normal mass ordering is assumed.}
\end{figure}

The simulation is in good agreement with the data, with a $\chi^2$/d.o.f.$=54.9/56$ for the energy-zenith angle histogram used in the fit, shown in Fig.\,\ref{fig:full_histogram}.  There, the zenith angle distribution of data and simulation is given for different energy bands. Below each histogram the ratio of data and the best-fit to the case of no-oscillations is also included. The maximum disappearance can be seen in the panel containing $E_\mathrm{reco} = [24 - 32]$\,GeV. The simulation agrees with the data at all energies considered. Figure\,\ref{fig:lefit} shows the agreement between data and simulation as a function of reconstructed baseline over energy $\left(L_\mathrm{reco}/E_\mathrm{reco}\right)$, a variable that does not directly enter the analysis. 

The 90\,\% confidence contours on the atmospheric oscillation parameters derived from this analysis, compared to the results from other experiments, are shown in Fig.\,\ref{fig:contours}. While this measurement is made at higher energies than other experiments (see Fig.\,\ref{fig:sample_energy}), the results are compatible and the precision achieved is comparable.

Higher statistics, ongoing improvements in veto algorithms, and the inclusion of cascade-like events will enhance the sensitivity of the oscillation studies with IceCube in the near future. This could be further improved by deploying additional instrumentation within the DeepCore array to collect more light per event and thus increase the statistics below $E_\nu=10$\,GeV, as proposed in the Precision IceCube Next Generation Upgrade (PINGU) \cite{pingu}.

\begin{acknowledgments}

We acknowledge the support from the following agencies:
U.S. National Science Foundation-Office of Polar Programs,
U.S. National Science Foundation-Physics Division,
University of Wisconsin Alumni Research Foundation,
the Grid Laboratory Of Wisconsin (GLOW) grid infrastructure at the University of Wisconsin - Madison, the Open Science Grid (OSG) grid infrastructure;
U.S. Department of Energy, and National Energy Research Scientific Computing Center,
the Louisiana Optical Network Initiative (LONI) grid computing resources;
Natural Sciences and Engineering Research Council of Canada,
WestGrid and Compute/Calcul Canada;
Swedish Research Council,
Swedish Polar Research Secretariat,
Swedish National Infrastructure for Computing (SNIC),
and Knut and Alice Wallenberg Foundation, Sweden;
German Ministry for Education and Research (BMBF),
Deutsche Forschungsgemeinschaft (DFG),
Helmholtz Alliance for Astroparticle Physics (HAP),
Research Department of Plasmas with Complex Interactions (Bochum), Germany;
Fund for Scientific Research (FNRS-FWO),
FWO Odysseus programme,
Flanders Institute to encourage scientific and technological research in industry (IWT),
Belgian Federal Science Policy Office (Belspo);
University of Oxford, United Kingdom;
Marsden Fund, New Zealand;
Australian Research Council;
Japan Society for Promotion of Science (JSPS);
the Swiss National Science Foundation (SNSF), Switzerland;
National Research Foundation of Korea (NRF);
Danish National Research Foundation, Denmark (DNRF)

\end{acknowledgments}

\end{document}